\documentclass{iaus}
\usepackage{graphicx}

\title[Intracluster PN] %% give here short title %%
{Intracluster Planetary Nebulae}

\author[Feldmeier]   %% give here short author list %%
{John J. Feldmeier$^1$}

\affiliation{$^1$Department of Physics and Astronomy, Youngstown State 
University, One University Plaza, Youngstown, OH, 44555 
USA \break email: jjfeldmeier@ysu.edu}

\pubyear{2006}
\volume{234}  %% insert here IAU Symposium No.
\pagerange{119--126}
\date{?? and in revised form ??}
\jname{Planetary Nebulae in Our Galaxy and Beyond - IAU Symposium 234}
\editors{M. J. Barlow \& R. H. Me\'ndez eds.}
\begin{document}

\maketitle

\begin{abstract}
I review the progress in research on intracluster planetary nebulae over 
the last five years.  Hundreds more intracluster planetary nebulae 
have been detected in the nearby Virgo and Fornax galaxy clusters, 
searches of several galaxy groups have been made, and intracluster
planetary candidates have been detected in the distant Coma cluster.  
The first theoretical studies of intracluster planetaries have 
also been completed, studying their utility as tracers of the 
intracluster light as a whole, and also as individual objects.  

From the results to date, it appears that intracluster planetaries 
are common in galaxy clusters (10-20\% of the total amount of starlight), 
but thus far, none have been detected in galaxy groups, a result 
which currently is not well understood.  Limited spectroscopic 
follow-up of intracluster planetaries in Virgo indicate that they 
have a complex velocity structure, in agreement with numerical 
models of intracluster light.  Hydrodynamic simulations of individual 
intracluster planetaries predict that their morphology is significantly 
altered by their intracluster environment, but their emission-line
properties appear to be unaffected.

\keywords{planetary nebulae: general, galaxies: interactions, 
galaxies: clusters}
\end{abstract}


\section{Introduction}

Intracluster starlight, the diffuse starlight which permeates many
galaxy clusters, is potentially of great interest to studies of galaxy 
and galaxy cluster evolution.  Since it is currently believed that the
bulk of the intracluster stars were originally formed within galaxies 
and then were tidally removed from them, they are an important way to 
study the mechanisms of tidal stripping and interactions that are common
within galaxy clusters (\cite[Dressler 1984]{dressler1984}).  
Modern numerical simulations of galaxy clusters show that the 
intracluster light is ubiquitous in galaxy clusters, has a complex spatial and
kinematic structure, and can be used to gain information on the 
dynamical evolution of galaxies and galaxy clusters (Napolitano et al. 2003; 
Willman et al. 2004; Murante et al. 2004; Sommer-Larsen et al. 2005; 
Stanghellini et al. 2006; Rudick et al. 2006).

However, the obstacles in observing intracluster light in detail are
substantial.  Due to its low surface brightness (at the brightest, 
less than 1\% of the night sky background in the $V$ band), 
it is extremely difficult to image directly.  There have been 
significant detections of intracluster light in galaxy clusters 
at low redshifts (z $<$ 0.3)
(Feldmeier et al. 2004a; Gonzalez et al. 2004; 
Mihos et al. 2005; Zibetti et al. 2005; Krick et al. 2006), 
but nearly all of these observations
require specialized observing techniques that are 
extremely time-consuming to carry out.  In addition, although 
intracluster imaging observations are crucial for obtaining a global 
view of the phenomenon, they can only give the spatial distribution of 
intracluster light, and possibly a color, meaning that detailed comparisons 
with the theoretical models will be difficult.  Finally, despite heroic
efforts, direct imaging cannot yet probe the lowest surface brightness 
features, which have surface brightnesses of $\mu_{V}$ = 32 mag/sq. arcsecond.

An alternate way to study intracluster light is to detect luminous 
individual intracluster stars in nearby galaxy clusters, and gain
more detailed information on the distribution, metallicity and velocities 
of intracluster stars than is possible from surface brightness
measurements.  This approach has also been quite successful: 
intracluster red giant stars (Ferguson, Tanvir, \& von Hippel 1998; 
Durrell et al. 2002), intracluster H~II regions 
(Lee et al. 2000; Gerhard et al. 2002; Ryan-Weber et al. 2004) 
and intracluster novae and supernovae (Gal-Yam et al. 2003; 
Neil, Shara, \& Oegerle 2005) have all been detected in galaxy clusters.

Here, we focus on another luminous tracer of the intracluster light: 
intracluster planetary nebulae (hereafter, IPN).  IPN have a number 
of unique advantages over other luminous tracers of the intracluster 
starlight.  Because IPN are emission-line objects, they can be detected 
efficiently in [O~III] $\lambda$ 5007 surveys from the ground.  
Therefore, using 
wide-field imagers common on 4-meter class telescopes, samples of 
hundreds of candidates can be found in a single telescope run.  With 
spectroscopic follow-up using 6-meter and larger telescopes, the radial 
velocities of IPN can be determined, offering the ability to study
the dynamics of intracluster starlight.  

\section{History of IPN research and Current Status}

The study of IPN is now over a decade old.  During a radial 
velocity survey of planetary nebulae (PN) 
candidates in the Virgo elliptical galaxy M~86, 
\cite{arna1996} found that 16 
of the 19 detected PN velocities were consistent with the galaxy's mean
velocity (v$_{radial}$ = -227 km s$^{-1}$).  The other three planetaries
had mean radial velocities of $\sim 1600$ km s$^{-1}$, more consistent with
the Virgo cluster's mean velocity.  \cite{arna1996} argued convincingly 
that these objects were intracluster planetary 
nebulae, and it is here that the term first enters the literature.  
Almost simultaneously, the first search for IPN candidates in
the Fornax cluster was published (Theuns \& Warren 1997), and more 
detections of IPN candidates in Virgo quickly followed 
(M\'endez et al. 1997; Ciardullo et al. 1998; Feldmeier, Ciardullo, 
\& Jacoby 1998).

However, a surprise was in the works.  Spectroscopic follow-up
of the IPN candidates revealed that some were not IPN, 
but instead background emission-line objects with extremely 
high equivalent width (Freeman et al. 2000; Kudritzki et al. 2000).  
This was unexpected,
because previous deep emission-line surveys had found very few
such objects (Pritchet 1994), though many have now been
detected at fainter magnitudes (Rhoads et al. 2004, and references therein).  
The most likely source 
of the contamination was found to be Lyman-$\alpha$ galaxies at
redshifts 3.12--3.14, where the Lyman-$\alpha$ $\lambda$ 1215 line
has been redshifted into the [O~III] $\lambda$ 5007 filters used
in IPN searches.  However other types of contaminating objects
may also exist (Stern et al. 2000; Norman et al. 2002).

Although these contaminants caused some consternation at first, 
a number of lines of evidence quickly showed
that the majority of IPN candidates are in fact, actual IPN.
Observations of blank control fields with identical search procedures
as the IPN surveys (Ciardullo et al. 2002; Castro-Rodr{\'{\i}}guez et al.
2003) have found that the contamination fraction is significant, but was 
less than the observed IPN surface density.  The surface densities 
found correspond to a contamination rate of $\approx$ 20\% in the 
Virgo cluster and $\approx$ 50\% in Fornax (Fornax is more distant
than Virgo, so its luminosity function is fainter, and therefore further down
the contaminating sources luminosity function).  There are still 
significant uncertainties in the background density due to 
large-scale structure, and to the small numbers of contaminating 
objects found thus far.  However, deeper and broader control fields 
are forthcoming (Gawiser et al. 2006).  Spectroscopic follow-up 
of IPN candidates (Freeman et al 2000; Ciardullo et al. 2002;
Arnaboldi et al. 2003, 2004), clearly show large numbers of
IPN candidates have the expected [O~III] $\lambda$ 5007 and 4959
emission lines, with a contamination rate similar to the blank
field surveys.  

Currently, with the widespread use of mosaic CCD detectors, and
automated detection methods derived from DAOPHOT and SExtractor,
over a hundred IPN candidates can be found in a single telescope
run (Okamura et al 2002; Arnaboldi et al. 2003; Feldmeier et al. 
2003, 2004b; Aguerri et al. 2005).  IPN candidates are readily identified 
as stellar sources that appear in a deep [O~III] $\lambda$ 5007 image, but 
completely disappear in an image through a filter that does not 
contain the [O~III] line.  Currently, over 400 IPN candidates have been 
detected in the Virgo cluster, and over 100 IPN candidates have been 
found in the Fornax cluster.  Figure~1 summarizes the status of the 
different imaging surveys.  However, only a small portion of 
these ($\approx$ 50) have any spectroscopic follow-up at all, and many 
of those spectra have low signal-to-noise.  However, despite all of 
the effort, to date, only a few percent of the total angular area of 
Virgo and Fornax have been surveyed.  Literally thousands of IPN wait 
to be discovered by 4-meter class telescopes.  

\begin{figure}
\caption{Images of portions of the Virgo cluster (left) and the
Fornax cluster(right).  The regions surveyed for IPN are shown as
the boxes in each image.  There are over 400 IPN candidates detected in
Virgo, and over 100 detected in Fornax at the current time.}\label{fig:wave}
\end{figure}

\section{Obtaining the amount of intracluster light from IPN}

In principle, determining the amount of intracluster luminosity from 
the observed numbers of IPN is straightforward.  Theories of simple 
stellar populations (Renzini \& Buzzoni 1986) have shown that the 
bolometric luminosity-specific stellar evolutionary flux 
of non-star-forming stellar populations should 
be $\sim 2 \times 10^{-11}$~stars-yr$^{-1}$-$L_{\odot}^{-1}$, (nearly) 
independent of population age or initial mass function.  If the 
lifetime of the planetary nebula stage is $\sim 25,000$~yr,
and if the empirical planetary nebula luminosity function (PNLF) 
is valid to $\sim 8$~mag below 
the PNLF cutoff, then every stellar system should have $\alpha \sim 50 \times 
10^{-8}$~PN-$L_{\odot}^{-1}$.  According to the empirical PNLF,
approximately one out of ten of these PNe will be within
2.5~mag of $M^*$.  Thus, under the above assumptions,
most stellar populations should have   
$\alpha_{2.5} \sim 50 \times 10^{-9}$~PN-$L_{\odot}^{-1}$.  
The observed number of IPN, coupled with the PNLF, 
can therefore be used to deduce the total luminosity
of the underlying stellar population.

In practice, there are a number of systematic effects that must be accounted
for before we can transform the numbers of IPN to a stellar luminosity
(Feldmeier et al. 2004b; Aguerri et al. 2005), which we briefly 
summarize here.  First, although stellar evolution theory originally 
predicted a constant $\alpha_{2.5}$ value for all non star-forming 
populations, observations (Ciardullo 1995; Ciardullo et al. 2005) and 
more sophisticated theoretical analysis (Buzzoni, Arnaboldi, \&
Corradi 2006) present a more complicated picture.   Since the 
amount of intracluster starlight derived is inversely proportional
to the $\alpha_{2.5}$ parameter, a large error in the amount of
intracluster light can result if this is not accounted for.  
By comparing the numbers of IPN in a field surrounding a 
{\sl HST} WFPC2~field, with RGB and AGB star counts, 
(Durrell et al. 2002) found a value of $\alpha_{2.5} = 23^{+10}_{-12} 
\times 10^{-9}$~PN-$L_{\odot}^{-1}$ for Virgo's intracluster population.  
Second, the IPN candidates of Virgo (and perhaps Fornax) may have a 
significant line-of-sight depth.  Since the conversion between 
number of PN and luminosity
depends on the shape of the luminosity function, this depth can
change the amount of intracluster light found from the data.
Models (Feldmeier, Ciardullo \& Jacoby 1998) indicate 
that the difference between simple models of the intracluster star
distribution, changes the derived intracluster star luminosity 
by up to a factor of three.  Thus far, all IPN researchers have 
adopted a single distance model in order to be conservative, but 
this effect is the least studied at this point.

After applying the corrections, the intracluster stellar
fractions for Virgo vary between 5 and 20\% (Feldmeier et al. 2004b; 
Aguerri et al 2005), with the errors being dominated by the 
systematic effects.  The IRG measurements (Ferguson, Tanvir \& 
von Hippel 1998; Durrell et al 2002) find somewhat less intracluster
light (10--15\%), but the various results also agree within the errors.

\section{What are the kinematics of the intracluster PN?}

Since most IPN candidates that are detected photometrically can be 
followed up spectroscopically, IPN are an ideal, and perhaps only, way 
for studying the kinematics of the intracluster light.  I will only 
briefly summarize this work here (see proceedings by Arnaboldi, Gerhard, 
this volume).  

The state of the art in IPN kinematics is the work by Arnaboldi et al. 
(2004).  Using a sample of 40 IPN over three fields in the Virgo cluster, 
Arnaboldi et al. (2004) showed that the radial velocity structure of
Virgo's IPN varies from field to field, and is often asymmetric, as was 
predicted by the numerical models of intracluster starlight.  This analysis
used less than 10\% of the known IPN candidates in Virgo: more spectroscopic 
follow-up is crucial for more detailed understanding.  Curiously, at least
three IPN of the Arnaboldi et al. (2004) have extreme velocities 
($\Delta v ~\approx$ 1000 km/s).  A possible explanation of these objects 
is given by Holley-Bockelmann et al. (2005) who proposes that some 
hypervelocity intracluster stars may be ejected by supermassive 
black hole binaries.

\section{What is the effect of the intracluster environment on the IPN?}

IPN are located in a fundamentally different environment than PN 
in our own Galaxy or in other galaxies.  They have enormous spatial
velocities (1000 - 2000 km s$^{-1}$) and are embedded 
within the hot (T = 10$^{6}$ - 10$^{7}$ K) gaseous intracluster medium.  
as they move within the confines of a galaxy cluster.  
What effect, if any, do these hostile conditions have on the structure 
and the emission properties of the nebula?  

At the the preceding IAU symposium, Feldmeier (2001) timidly suggested 
that the IPN might become aspherical, and possibly fragment, due to 
Rayleigh-Taylor instabilities.  Recent theoretical work by 
Villaver \& Stanghellini (2005; see also Villaver, this volume) 
has added immensely to our understanding of the internal structure of IPN.  
Using a hydrodynamic simulation of a 1 M$\odot$ 
main sequence star progenitor, and physical conditions similar to a IPN 
within the Virgo cluster, Villaver \& Stanghellini (2005) have found that
the IPN becomes very aspherical in appearance compared to normal 
planetaries, with a long ($\approx 130$~pc) gas stream that trails behind the 
IPN's direction of motion.  However, Villaver \& Stanghellini (2005) 
also find that the recombination emission appears to be unaffected
in these objects, implying that the [O~III] $\lambda$ 5007 emission 
is unaffected as well.  An interesting open question is whether any 
signs of the strong shocks seen in the simulations could be observed 
in a deep spectrum of the IPN.

\section{IPN in more distant clusters}

An exciting recent development is the detection of IPN 
candidates in the Coma cluster of galaxies, an extremely rich 
cluster at $\approx$ 100 Mpc, over a
factor of six more distant than all previous IPN searches 
(Gerhard et al. 2004; see also Gerhard, Arnaboldi, this volume)
These candidates were found using a technique of filling the entire 
focal plane of the 8-m Subaru telescope with a slit mask through 
the appropriate [O~III] $\lambda$ 5007 filter, and looking for narrow-line 
emission sources.  Deep broad-band imaging has implied that the Coma Cluster 
may have a very high intracluster star fraction, up to 50\% 
(Bernstein et al. 2005), so it is 
quite plausible that a few IPN could be detected, 
despite the relatively small area surveyed.  The most exciting aspect 
of this observation is that it opens up a way to observe PN at 
much greater distances than previously thought possible, and 
makes it possible to place IPN density limits in more distant 
galaxy clusters.  Additional observations using this method are now
underway (Gerhard, private communication)     

\section{IPN searches in Galaxy Groups}

In contrast, searches for IPN in galaxy groups have been less fruitful.
Searches have been made of the M~81 galaxy group (Feldmeier et al. 2001;
Feldmeier et al. 2006, in prep.), the Leo~I galaxy group 
(Castro-Rodriguez et al. 2003), and the Hickson compact group 
HCG~44 (Castro-Rodriquez et al. 2005).  In all three of these searches, 
no genuine IPN candidates have yet been found.  Some emission-line sources 
have been detected, but their properties are consistent with background 
objects.  The non-detections place a strong limit on intra-group starlight
in these groups, over the regions searched, to limits of a few percent.

When compared with other measurements of intracluster star fractions    
through modern deep imaging (Feldmeier et al. 2004; Gonzalez et al. 2004; 
Zibetti et al. 2005; Krick et al. 2006), or through the detection of 
intracluster supernovae (Gal-Yam et al. 2003), an interesting pattern 
emerges, plotted in Figure~2.  For galaxy clusters, the data is 
consistent with an approximate mean fraction of 15--20\%.  When we 
move to the group environment, the fraction abruptly drops, with no 
sign of any smooth decline.  This feature, which we have 
dubbed the ``intracluster cliff,'' is currently unexplained.

However, a theoretical paper by Sommer-Larsen (2006) has 
claimed that the intra-group fractions are quite high, between 12 and 45\%
How can this discrepancy be explained?   Villaver \& Stanghellini (2005) 
suggest that the higher density of the intragroup environment might cause 
a larger amount of gaseous stripping, and therefore the intra-group PN
might be undetectable.  Sommer-Larsen (2006) claims that the intra-group PN
would be widely scattered, and the surveys to date do not cover a broad 
enough spatial range to detect the few objects that would be expected.  
Clearly, more observational studies are needed, over larger spatial scales 
of nearby galaxy groups to strengthen these results.  An IPN survey of the 
intermediate Ursa Major galaxy cluster has also begun to address 
these questions.

\begin{figure}
\includegraphics[width=6in]{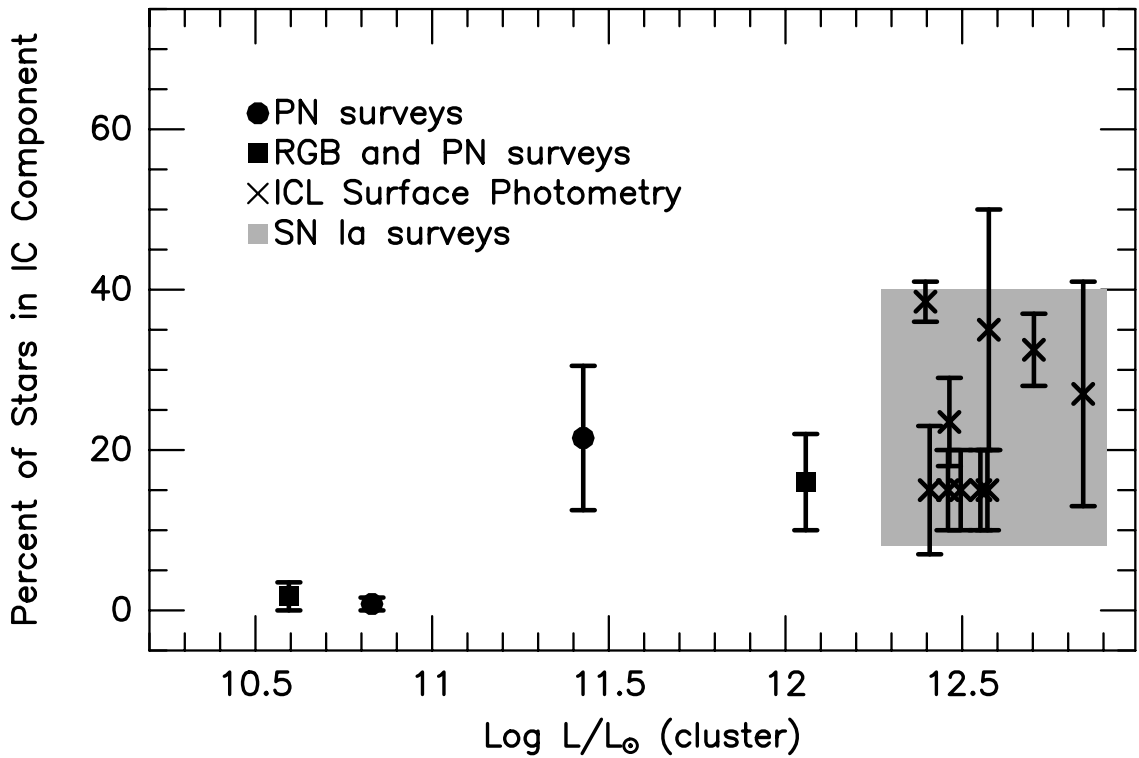}
\caption{The fraction of intracluster starlight detected as
a function of the cluster's total $B$-band luminosity.  
The data are consistent with a intracluster fraction 
of $\sim$ 20\% for massive 
systems, and then a drop to effectively zero for the two lowest mass
groups.  This ``intracluster cliff'' implies that something in the
cluster environment promotes intracluster star production, but more
data is needed to confirm these results.}\label{fig:cliff}
\end{figure}

\begin{acknowledgments}
I would like to thank the conference organizers for allowing me to give
this review, and for running an excellent conference.  I would also 
like to thank my collaborators for their years of effort on the science 
presented here.  I would also like to thank Magda Arnaboldi, 
Ortwin Gerhard and Eva Villaver for sharing unpublished results 
previous to the conference.  This work was supported in part by 
NSF grant AST 0302030.
\end{acknowledgments}

\end{document}